\documentclass[prl, superscriptaddress, twocolumn, showpacs, aps, floatfix, nofootinbib]{revtex4-1}

\usepackage{amsfonts}
\usepackage{amssymb}
\usepackage{graphicx}
\usepackage{color}
\usepackage{amsmath}
\usepackage[bookmarksnumbered, bookmarks, breaklinks, linktocpage]{hyperref}

\newcommand{\Tr}        {\mathrm{Tr}}

\newcommand{\ket}[1]    {| #1 \rangle}
\newcommand{\bk}[2]     {\langle #1 | #2 \rangle}
\newcommand{\kb}[2]     {| #1 \rangle \! \langle #2 |}
\newcommand{\cH}        {{\mathcal H}}
\newcommand{\cS}        {{\mathcal S}}
\newcommand{\cA}        {{\mathcal A}}
\newcommand{\cE}        {{\mathcal E}}

\newcommand{\eend}      {\hspace{\stretch{1}}\rule{1ex}{1ex}}

\newcommand\dpcom[1]{}

\begin{document}

\title{Wavepacket collapse and the core quantum postulates: Discreteness of \\ quantum jumps from unitarity, repeatability, and actionable information}

\author{Wojciech H. Zurek}
\affiliation{Theory Division, LANL, MS B213, Los Alamos, NM  87545}
\affiliation{Santa Fe Institute, 1399 Hyde Park Rd.,  Santa Fe, NM 87501 USA}

\date{\today}

\begin{abstract} Unknown quantum state of a single system cannot be discovered as a measured system is re-prepared -- it jumps into an eigenstate of the measured observable. This impossibility to find out a quantum state and other symptoms usually blamed on the wavepacket collapse follow (as was recently demonstrated for pure states of measured systems) from unitarity (that does not, however, allow for a {\it literal} collapse) and from the repeatability of measurements:  Continuous unitary evolution and repeatability suffice to establish discreteness that underlies quantum jumps. Here we consider mixed states of a {\it macroscopic}, open system (such as an apparatus), and we allow its {\it microscopic} state to change when, e.g., measured by an observer, providing that its salient features remain unchanged -- providing that observers regards it as the same record. We conclude that repeatably accessible states of macroscopic systems (such as the states of the apparatus pointer) must correspond to orthogonal subspaces in the Hilbert space. The symmetry breaking we exhibit defies the egalitarian quantum superposition principle and unitary symmetry of the Hilbert space, as it singles out preferred subspaces. We conclude that the resulting discreteness (that underlies quantum jumps) emerges from the continuity of the core quantum postulates plus repeatability also in macroscopic and open -- but ultimately quantum -- systems such as measuring devices accessed by observers, where (in contract to pure states of microsystems) repeatability is paramount.
\end{abstract}

\maketitle

\section{Introduction} 

Repeatability of measurements -- the demand that immediate re-measurement should give the same outcome -- is a textbook postulate of quantum theory \cite{Dirac}. It is an extreme example of predictability that motivates the very idea of a state: States are useful because they allow for predictions. Repeatability is the simplest case. It allows one to confirm system is in a state in which it is known to be. In quantum theory, where usual attributes of existence of states -- e.g., the ability to find out what they are -- are severely constrained, repeatability ties their role as a summary of information \cite{Bohr} and their function as building blocks of reality: Repeatability allows one to confirm that what is known to exist is indeed ``out there''.

Repeatability is also a {\it necessary} symptom of the wave-packet collapse. Quantum states are simultaneously determined and prepared by measurements. The long-lived \cite{Bohr,Bohr2,vonN,Everett,Everett2,WZ,Auletta} but still lively debate about quantum measurements became, with the advent of decoherence and einselection \cite{Zurek91, Zurek03b,Schlosshauer,GJK+96a}, focused on symptoms of collapse one can reproduce using only unitary evolutions, and whether repeatability and predictability are {\it sufficient} to account for perception of definite outcomes and classical reality. 

Recent progress includes demonstration \cite{Zurek07} that only orthogonal states of a system $\cS$, 
$\bk u v=0$, 
can be repeatably measured and remain unchanged 
while they leave distinct (even if imperfect) imprints on other quantum systems (e.g., an apparatus $\cA$ or an environment $\cE$).
Thus, when information transfer from $\cS$ to $\cA$ is unitary;  
$$
\ket u \ket {A_0} \longrightarrow \ket u \ket {A_u}, \ \ \ \  \ket v \ket {A_0} \longrightarrow \ket v \ket {A_v} \eqno(1)
$$ 
and leaves states $ \ket u,~\ket v$ of $\cS$ untouched (guaranteeing repeatability) preservation of scalar products:
$$
\bk u v = \bk u v \bk {A_u} {A_v} \eqno(2)
$$
follows. When $\bk u v \neq 0$ this implies $ \bk {A_u} {A_v} = 1$. Consequently, $\ket {A_u} $ and $ \ket {A_v} $ can differ  
only when $\bk u v = 0$ -- that is, when dividing both sides of Eq.~(2) by $\bk u v$ is illegal. So, extracting information that distinguishes $\ket u$ from $\ket v$ is possible only when $\ket u$ and $\ket v$ are orthogonal.

For pure states this accounts for the ``business end" of the collapse postulate: Measurements that respect unitarity cannot reveal unknown states of the system. Thus, sets of quantum states that can exist in a sense that approximates classical existence -- states that can survive multiple confirmations of their identity -- must be orthogonal. Their superpositions are perturbed when measured by $\cA$, as $\alpha \ket u + \beta \ket v$ will entangle with the apparatus:
$$
(\alpha \ket u + \beta \ket v)\ket {A_0} \rightarrow  \alpha \ket u \ket {A_u}  + \beta \ket v \ket {A_v}.
$$
Equation (2) also justifies 
(using quantum ingredients) textbook demand for the observables to be Hermitean. Generalization 
to when systems that receive information (above, $\cA$) are mixed is possible \cite{Zurek07,Z07}. This allows mixed states of $\cA$ or an environment, as $\cE$ plays a role of $\cA$ in decoherence. 

Orthogonality implied by Eq. (2) signals emergence of {\it pointer states} \cite{Zurek81,Zurek82}. They appear because information transfer is in a sense nonlinear: To find out a state, as in cloning \cite{WZ82,Dieks82,Yuen}, one demands a copy of the preexisting state of $\cS$. Tension between this nonlinear demand and linearity of quantum evolutions can be reconciled only for selected orthogonal sets of states. Information transfer leads to preferred basis that breaks unitary symmetry of the Hilbert space $\cH_{\cS}$. 
Superpositions of preferred states are still legal: They can persist in isolation, but they cannot be found out by the same measuring device. Existence (persistence attested by many copies in the records kept by measuring devices or environments) is recovered at the price of unitary symmetry breaking that singles out states which survive measurements.

In decoherence theory \cite{Zurek91,Zurek03b,Schlosshauer,GJK+96a} the ability to withstand scrutiny of the environment defines preferred, einselected pointer states \cite{Zurek81,Zurek82}. Their resilience is crucial in quantum Darwinism \cite{OPZ04,BKZ06,Z09,PazRoncaglia,ZQZ10,RZ11}. Quantum Darwinism derives objective reality from multiple copies of the information about preferred states of the system deposited in the environment by decoherence. The ability of the originals to survive copying is then essential. As we have seen, emergence of such preferred states can be deduced without delving into the structure of Hamiltonians (which is the usual approach). 

\section{Repeatability of copies and the role of degeneracy of the original}

Repeatability is a special case of predictability. Predictability sieve \cite{ZHP,Zurek93} selects pointer states based on their immunity to interactions with $\cE$. 
We shall study the case when the copied system can be perturbed, but the changes of the original state do not affect copies. This can happen in presence of degeneracy, i.e., when different original microstates of $\cS$ produce the same copy with only partial (coarse-grained) information about it. 

To motivate our study we note that measurements on microscopic systems are rarely  repeatable. A state of the measured system is usually altered, and some systems disappear completely (photons are usually absorbed when detected). Where measurements of quantum systems are concerned, exceptions (i.e., ``nondemolition measurements'') to this ``mistreatment'' are rare, approximate, and generally very expensive to implement.

By contrast, {\it records} -- states of memory -- are faithfully preserved. For instance, apparatus pointer can be consulted and read off by many. Similar resilience under scrutiny characterizes classical systems. 
Yet, memory of a measuring device, or, indeed, any effectively classical system is made out of quantum components, so it is ultimately quantum. Therefore, states of such systems (including records of measurement outcomes carried out using macroscopic apparatus) are, ultimately, quantum. Thus, the ease of finding such states -- of copying memory content without worries about repeatability (i.e., reliability of the records) and without nondemolition measurements -- may be surprising. 
However -- and this is the key hint to how repeatability \cite{Dirac} can be relaxed -- the {\it complete} microscopic state of the memory device need not be copied or even faithfully preserved by the readout. Indeed, memory devices are immersed in their environments, and their microscopic states are largely ignored by readout, i.e., by the copying. Only a small subset of degrees of freedom that contain the record is accessed. 

In practice repeatability is established by comparing copies. Records they reveal (the copies {\it per se}) should tell the same story. Thus, when $\cS$ is macroscopic -- a memory device such as a RAM or an apparatus pointer -- instead of Eq. (1) one should consider a system in a mixed state and a series ($a,~b,~c...$) of copying transitions;
$$ 
\ \ \ \ \rho_{\cS }^u \kb {A_0} {A_0} \kb {A'_0} {A'_0} ... \stackrel {a} {\longrightarrow} \tilde \rho_{\cS }^u \kb {A_u} {A_u} \kb {A'_0} {A'_0}  ...\ , \eqno(3u)
$$ 
\vspace{-0.7cm}
$$
\ \ \ \tilde \rho_{\cS }^u \kb {A_u} {A_u} \kb {A'_0} {A'_0}... \stackrel {b} {\longrightarrow} {\tilde {\tilde {\rho}}}_{\cS }^u \kb {A_u} {A_u} \kb {A'_u} {A'_u}...\ .  
$$
Repeatability means repetition of records -- the same subscript labeling state of recording devices $\cA$, $\cA'$, etc. This allows differences between $\rho_{\cS }^u $, $\tilde \rho_{\cS }^u$, ${\tilde {\tilde {\rho}}}_{\cS }^u$,... We only demand that the property of $\cS$ that affects states of $\cA$, $\cA'$, etc,. be unchanged by the acts $a,~b,~c,... $ of copying.

Such identity of copies 
can arise in absence of the sameness of originals, as is seen in a pure state example:
$$
\ket {u_1} \ket {A_0} \ket {A_0} \stackrel {a} {\longrightarrow} \ket {\tilde u_1} \ket {A_u} \ket {A_0} \stackrel {b} {\longrightarrow} \ket {\tilde { \tilde {u}}_1} \ket {A_u} \ket {A_u} ; 
$$
\vspace{-0.7cm}
$$
\ket {u_2} \ket {A_0} \ket {A_0}  \stackrel {a} {\longrightarrow} \ket {\tilde u_2} \ket {A_u} \ket {A_0} \stackrel {b} {\longrightarrow} \ket {\tilde { \tilde {u}}_2} \ket {A_u} \ket {A_u} . \eqno(4)
$$
This can happen when $\ket {u_1}$ and $\ket {u_2}$ (and their ``descendants'') belong to the same degenerate subspace {\bf u} of the observable measured on $\cS$. Linearity implies that their superpositions will also deposit the same record in $\cA$:
$$
(\alpha \ket {u_1} + \beta \ket {u_2} ) \ket {A_0} \ket {A_0'} \stackrel {a} {\longrightarrow} ... \stackrel {b} {\longrightarrow}(\alpha \ket {\tilde { \tilde {u}}_1}+\beta \ket {\tilde { \tilde {u}}_2} ) \ket {A_u} \ket {A_u'} \eqno(5) 
$$
Thus, superpositions of states that belong to eigenspace $\bf u$ of the observable controlling evolution of $\cA$ will result in the same record. States that leave different imprints on $\cA$, $\cA'$, etc.,  have to -- by reasoning that parallels Eq. (2) -- belong to a different, orthogonal subspace, e.g., $\bf v \perp \bf u$. Therefore, measured observable can be expressed as:
\vspace{-0.2cm}
$$ \hat {\bf o} = \sum_k o_k {\bf P}_k  \eqno (6)$$
We conclude this ``pure degeneracy'' case by noting that eigenspaces ${\bf P}_k$ (such as $\bf u,~v$ above) that induce different imprints on $\cA$ while the underlying states of $\cS$ remain confined inside them must be orthogonal. This constitutes a significant extension of the consequences of Eqs. (1),~(2). 

We now demonstrate that a similar reasoning applies to mixed states of a macroscopic $\cS$. That is, {\it we show that mixed states can produce distinguishable copies of the records they hold only when they have support in non-overlapping -- orthogonal -- subspaces of the Hilbert space of $\cS$.}

{\bf Proof:} To extend the proof of orthogonality of states that survive copying intact to mixtures we write down ``$v$'' counterpart of Eq. (3$u$):
$$\ \ \ \ \ \rho_{\cS}^v \kb {A_0} {A_0} \kb {A'_0} {A'_0} ... \stackrel {a} {\longrightarrow} \tilde \rho_{\cS }^v \kb {A_v} {A_v} \kb {A'_0} {A'_0}  ... \ ,  \eqno(3v)
$$ 
\vspace{-0.9cm}
$$
\ \ \ \ \tilde \rho_{\cS }^v \kb {A_v} {A_v} \kb {A'_0} {A'_0} ...  \stackrel {b} {\longrightarrow}{\tilde {\tilde {\rho}}}_{\cS }^v \kb {A_v} {A_v} \kb {A'_v} {A'_v} ... \ .
$$
The reasoning used for pure states can be repeated. Unitarity implies preservation of the Schmidt-Hilbert norm:
$$\Tr \rho_{\cS}^u \rho_{\cS }^v = \Tr \tilde \rho_{\cS}^u \tilde  \rho_{\cS }^v ~ |\bk {A_u} {A_v}|^2 \eqno(7a) $$
\vspace{-0.9cm}
$$ \Tr \rho_{\cS}^u \rho_{\cS }^v = ~ \Tr {\tilde {\tilde \rho}}_{\cS}^u {\tilde {\tilde \rho}}_{\cS }^v |\bk {A_u} {A_v}|^4 \eqno(7b)
$$
Moreover, when states $\ket {A_u}, \ket {A_v}$  that ``tag'' state of $\cS$ are pure, $\Tr \rho_{\cS}^u \rho_{\cS}^v = \Tr \tilde \rho_{\cS}^u \tilde  \rho_{\cS }^v=...$ follows -- unitary evolution that preserves purity of tags also preserves eigenvalues of density matrices $\rho_{\cS}^u$, $\rho_{\cS}^v$. Hence, eigenstates $\ket {u_k}$ ($\ket {v_k}$) of $\rho_{\cS}^u$ ($\rho_{\cS}^v$) that span $\bf u$ ($\bf v$) evolve as dictated by Eqs.~(4,~5):
$$
\ket {u_k} \ket {A_0} \longrightarrow \ket {\tilde u_k} \ket {A_u} , \ \ \ 
\ket {v_k} \ket {A_0} \longrightarrow \ket {\tilde v_k} \ket {A_v} 
$$
Equality of Schmidt-Hilbert norms  before and after tagging implies that
distinguishable tags ($ |\bk {A_u} {A_v}|^2 < 1$) can be imprinted on $\cA$ only when 
$$\Tr \rho_{\cS}^u \rho_{\cS }^v = \Tr \tilde \rho_{\cS}^u \tilde  \rho_{\cS }^v
=\Tr {\tilde {\tilde \rho}}_{u\cS} {\tilde {\tilde \rho}}_{v\cS }
=...= 0 . $$
Otherwise, $ |\bk {A_u} {A_v}|^2=1$ -- no information about $\cS$ is acquired by $\cA$, and the measurement is a failure. QED.

The assumption that suffices to assure repeatability (of copies) is very much weaker than the obvious demand that the mixed state of the system be essentially unchanged. We have tested it assuming that the state of $\cA$'s that contain consecutive copies are pure. We shall now see that even mixed states of $\cA$'s are sufficient, but we will need to introduce a test that allows one to confirm that they contain the same essential information that was obtained from $\cS$.

\section{Actionable information, mixtures, and quantum jumps}

The key idealization so far was the purity of the states of $\cA$. We now relax it and allow $\cA$ not only to be mixed, but also to interact with a decohering environment $\cE$, and retain correlations with the other systems (including $\cS$ and $\cE$):
$$ \rho_{\cS}^u \rho_{0\cA} \rho_{0\cA '} ...\rho_{0\cA^{(k)}}...  \rho_{0\cE} \longrightarrow \varrho^u_{ \cS \cA \cA' ... \cA^{(k)} ... \cE} \ , \eqno(8u)$$
\vspace{-0.7cm}
$$ \rho_{\cS}^v \rho_{0\cA} \rho_{0\cA '} ... \rho_{0\cA^{(k)}} ... \rho_{0\cE} \longrightarrow \varrho^v_{\cS \cA \cA' ... \cA^{(k)} ... \cE} \ . \eqno(8v) $$
The composite state $\varrho_{\cS \cA \cA' ... \cA^{(k)} ... \cE}^u$ obviously contains information distinguishing it from $\varrho_{\cS \cA \cA' ... \cA^{(k)} ... \cE}^v$, but we want it contained in every $\cA^{(k)}$ and check if it is reflected in the reduced $\rho_{\cA^{(k)}}^u $ obtained from $\varrho_{ \cS \cA \cA' ... \cA^{(k)} ... \cE}^u$. To verify this we devise an operational test: Information distinguishing $u$ from $v$ is present in $\cA^{(k)}$ when it is it can be acted upon -- when it is {\it actionable}, i.e., when there exists a unitary ${\cal U}_{{\cA^{(k)}}{\cal T}}$ coupling $\cA^{(k)}$ with a test system ${\cal T}$ that can induce conditional dynamics -- transform its state from initial $\tau_0$ to record-dependent and distinct $\tau_u$ or $\tau_v$:
$$\varrho_{ \cS \cA \cA' ... \cA^{(k)} ... \cE}^u \otimes \tau_0 \stackrel {{\cal U}_{{\cA^{(k)}}{\cal T}}} \longrightarrow \tilde \varrho_{\cS \cA \cA' ... \cA^{(k)} ... \cE}^u \otimes \tau_u \ , \eqno(9u)$$
\vspace{-0.7cm}
$$ \varrho_{\cS \cA \cA' ... \cA^{(k)} ... \cE}^v \otimes \tau_0 \stackrel {{\cal U}_{{\cA^{(k)}}{\cal T}}} \longrightarrow \tilde \varrho_{ \cS \cA \cA' ... \cA^{(k)} ... \cE}^v \otimes \tau_v \ . \eqno(9v)
$$
Copying of the information that distinguishes between $u$ and $v$ from $\cA^{(k)}$ to ${\cal T}$ would be an example that proves $\cA^{(k)}$ contains actionable information about $\cS$.

{\bf Proof}: The state of $\cal T$ need not be pure. However, density matrices $\tau_0,~\tau_u,~\tau_v$ should form a product state with other systems (as emphasized by notation), so that while the eigenstates of the density matrix representing $\cal T$ rotate conditionally in response to the record in $\cA^{(k)}$, its eigenvalues remain unchanged.
That is, $\rho^u_{\cA^{(k)}} $ and $\rho^v_{\cA^{(k)}} $ should act as two logical states of the control qubit in a $\tt cnot$, while $\cal T$ acts as target qubit. It is now straightforward to see (using Eqs. (8), (9), and preservation of Schmidt-Hilbert norm under ${\cal U}_{{\cA^{(k)}}{\cal T}}$) that as long as $\tau_u$ differs from $\tau_v$, i.e. $\Tr ~ \tau_u \tau_v < \Tr  ~\tau_0^2$, we necessarily have:
$ \Tr ~\rho_{\cS}^u  \rho_{\cS}^v (\rho_{0\cA} \rho_{0\cA '} ...\rho_{0\cA^{(k)}}...  \rho_{0\cE})^2 = \Tr ~ \varrho^u_{ \cS \cA \cA' ... \cA^{(k)} ... \cE} \varrho^v_{ \cS \cA \cA' ... \cA^{(k)} ... \cE} = 0$, which immediately implies orthogonality of the record states of $\cS$:
$$ \Tr ~\rho^u_{\cS}  \rho^v_{\cS}=0 \eqno(10)$$
for mixed, decohering $\cA$'s. QED.

We have now established that actionable information about $\cS$ can be repeatedly passed on to other quantum systems $\cA^{(k)}$ only when it concerns orthogonal subspaces of $\cS$. This is our main result. It is a vast extension of Ref.  \cite{Zurek07}, relevant for mixtures of $\cS$ that can be affected by measurement or decoherence. It implies discreteness (the reason for quantum jumps) in realistic settings where macroscopic (but quantum) $\cS$ holds records of states $u$ and $v$ of e.g. a measured quantum microsystem. 

A few comments are in order. We first note that actionable information we have tested for above is local: It resides in a specific $\cA^{(k)}$. This is assured by the conditional evolution operator ${\cal U}_{{\cA^{(k)}}{\cal T}}$ that couples only $\cA^{(k)}$ to $\cal T$. This need not be always the case: Actionable information may reside in nonlocal correlations between systems. We shall see such an example below. 

Definition of actionability can be extended by allowing non-unitary evolutions implemented with the help of ancillae and/or environments. Care must be taken, however, with such extensions: When actionability is meant to test if the information resides in $\cA^{(k)}$, one must assure that it is not supplied by $\cE$, ancilla, or the correlations with or between them. To ascertain this one should start such additional systems in a product state that has no record of the outcomes $u,v$, etc., as was done with $\cE$ on the RHS of Eq. (8). 

One interesting generalization is the possibility of very different $\cA$'s (e.g., computer memory, piece of paper, etc.) in which copies are stored. Actionability can be still defined by choosing a $k$-dependent unitary ${\cal U}^{(k)}_{{\cA^{(k)}}{\cal T}}$ that suites the nature of the specific ${\cA^{(k)}}$. 

Our assumptions above result in discreteness -- they suffice to prove orthogonality of record states of $\cS$. They are not overwhelmingly strong -- all we have asked for was existence of {\it some} unitary that can cause a suitable conditional evolution of {\it some} state of $\cal T$. Still, it is interesting to enquire whether these assumptions can be further relaxed \cite{foot}. Note that while mixed states of $\cA$ and $\cal T$ are allowed, product structure of the final state, Eq. (9) (rather than purity of $\cal A$ or $\cal T$) was key for the proof \cite{note}. 

\section{Actionable mixtures don't mix}

We have just established that actionable records -- information that can be acted upon -- reside in orthogonal subspaces of Hilbert space. An intriguing followup concerns the possibility of using linear combinations of record states as records of actionable information. 

The motivation for this question comes from considering pure states. As we have seen (already in the introduction) pure states need to be orthogonal in order to act as originals for (even imperfect) copies. However, one could also use as originals their linear combinations -- their superpositions -- as long as they are orthogonal. For instance, $\ket 0$ and $\ket 1$ can act as logical states in the control qubit of {\tt cnot}, but it is also possible to use states $\ket +$ and $\ket -$ -- their superpositions $\ket {\pm} = (\ket 0 \pm \ket 1)/\sqrt 2$ in a (different) controlled-not. 

The question arises whether one can play a similar trick with mixed states that constitute actionable records -- i.e., use their linear combinations as actionable records of something else. The answer is, perhaps surprisingly, that this is impossible. To see why, consider $\rho_u$ and $\rho_v$ that are actionable. We can then show that,
{\it when there is a unitary that transforms a test system $\tau$ depending on whether $\cal S$ is in the state $\rho_u$ or $\rho_v$, then there is no unitary transformation that can be simiarily conditioned on linear combinations of  $\rho_u$ and $\rho_v$.}

{\bf Proof}: Suppose a transformation ${\cal U}'$ that can be conditioned on:
$\rho_{ab} =  a \rho_u + b \rho_v$ and $\rho_{cd} =  c \rho_u + d \rho_v$ (where $a,~b,~c,~d$ are non-negative and $a+b=1, \ c+d=1$) exists. Then there is a test system $\varsigma$ that can have its state altered depending on the content of the ``mixed'' records as:
$$ (a \rho_u + b \rho_v) \sigma_0  \stackrel  {{\cal U}'} \longrightarrow  (a \tilde \rho_u + b \tilde \rho_v) \sigma_{ab} $$
\vspace{-0.7cm}
$$ (c \rho_u + d \rho_v) \sigma_0  \stackrel  {{\cal U}'} \longrightarrow  (c \tilde \rho_u + d \tilde \rho_v) \sigma_{cd} $$
Reasoning as earlier we obtain that $\Tr \sigma_0^2 \neq \Tr \sigma_{ab} \sigma_{cd}$ iff:
$$\Tr (a \rho_u + b \rho_v)(c \rho_u + d \rho_v)=\Tr (a \tilde \rho_u + b \tilde \rho_v)(c \tilde \rho_u + d \tilde \rho_v) = 0$$
However, by assumption $\rho_u$ and $\rho_v$ constitute actionable records. Therefore, $\rho_u \perp \rho_v$ and $\tilde \rho_u \perp \tilde \rho_v$. Consequently, the equality above leads one to conclude that:
$$ \Tr (ac \rho_u^2 + bd \rho_v^2) = \Tr (ac \tilde \rho_u^2 + bd \tilde \rho_v^2) = 0$$
which immediately implies that either $a=0, ~b=1$, and therefore $d=0,~c=1$, or, alternatively, $b=0, ~a=1$ and therefore $c=0, ~d=1$. Therefore, indeed, non-trivial combinations of actionable mixtures cannot be actionable. QED.

This is a surprising result. Actionable mixtures cannot mix and remain actionable. The difference with the case of pure states was that there the coefficients (analogues of $a,~b,~c,~d$ could be negative (or complex) so their linear combinations could be orthogonal.

We note that the above theorem does not preclude coarse-graining (or fine-graining). That is, several actionable mixtures can be combined as a single actionable state in arbitrary proportions, providing that they are now regarded as a single record.

\section{Repeatability and POVM's}

We have seen that only orthogonal states (of $\cS$) can act as originals for an unlimited numbers of copies (in $\cA$'s). However, it is also well known that many outcome states detected in actual measurements are not orthogonal. Measurements that result in such outcomes are not repeatable -- there is no contradiction -- but it is important to see how this ``fact of life'' fits into our discussion.
This realization also suggests a simple but nevertheless nontrivial connection between distinguishability and repeatability, illustrated in Fig. 1.

{\begin{figure}[tb]
\begin{tabular}{l}
\vspace{-0.15in} 
\includegraphics[width=3.5in]{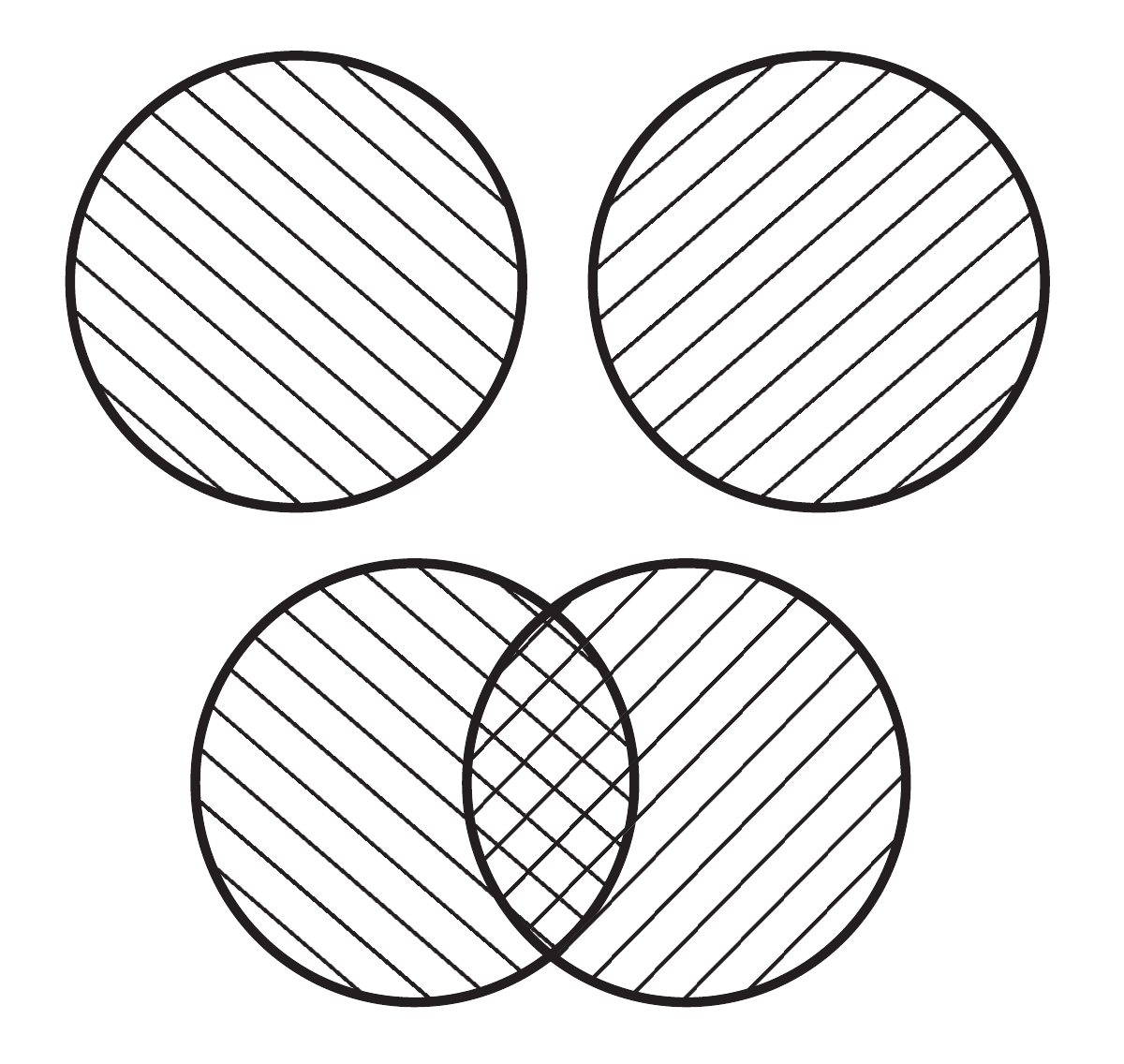}\\
\end{tabular}
\caption{A pre-quantum, fundamental interdependence of distinguishability and repeatability: The two circles correspond to two outcomes -- two properties of the underlying state
(represented by two different cross-hatchings). 
A measurement that can result in either outcome -- produce a record correlated with either of these two states -- can be repeatable only when the two properties are mutually exclusive (the the two states -- two circles) do not overlap (case illustrated at the top). Repeatability is impossible without distinguishability: When two outcomes are not mutually exclusive, states overlap (case illustrated in the bottom), so repetition of the measurement can always result in a system switching the state (and, thus, defying repeatability).
In the quantum case (discussed in this paper) this pre-quantum interdependence of repeatability and distinguishability ultimately leads to discreteness of the underlying quantum jumps -- to the orthogonality of repeatable measurement outcomes. 
However, the basic intuition demanding distinguishability as a prerequisite of repeatability does not rely on quantum formalism.}
\label{Distinguishability}
\end{figure}

The answer is simple, and we have anticipated it earlier: Only {\it records} of outcomes need to be orthogonal, as only they -- and not the original quantum states of the microsystem -- need to be repeatedly accessible. This suggests how positive operator valued measures (POVM's) that do not correspond to orthogonal outcomes arise: It is natural to define POVM using non-commuting Hermitian  observables \cite{Helstrom}. A simple example is a measurement of non-commuting $\hat Y$ and $\hat Z$ with the eigenstates $\{ \ket {y_k} \}$ and $\{ \ket {z_l} \}$. The measurement of $\hat Y$ followed by a measurement of $\hat Z$ is associated with a pair of numbers $k,~l$ that correspond to outcomes $\ket {y_k}$ and $\ket {z_l}$. The operator:
$$ F^{Y,Z}(k,l)=\kb {y_k}{y_k} U^\dagger_t \kb {z_l}{z_l} U_t \kb {y_k}{y_k} 
$$
describes the effect of the measurement of $\hat Y$ (outcome $k$) followed by the free evolution  $U_t$ followed by the measurement of $\hat Z$ (outcome $l$) on the system. It is clear that $F^{Y,Z}(k,l)$ are Hermitian and non-negative definite, but -- unless $[\hat Y, U^\dagger_t \hat Z U_t]=0$ -- they are not projectors. Indeed, there are too many $F^{Y,Z}(k,l)$ to be orthogonal projectors in $\cal H_S$. Yet, they resolve the identity operator:
$ \sum_k\sum_l F^{Y,Z}(k,l) = {\bf 1}_S$.
The probability of record $(k, l)$ is
$ p_{k,l}=\Tr F^{Y,Z}(k,l) \rho(0)$.
This example shows how POVM's can arise alongside the repeatability of records. Moreover, this is a realistic example: Monitoring of the same observable (such as position) of an evolving system such as a harmonic oscillator is covered by it, as, in general, $[\hat X, U^\dagger_t \hat X U_t] \neq 0$. Measurements of non-commuting observables ($\hat X$ and $\hat P$) 
can be treated similarly.

\section{Repeatability in the (church of) larger Hilbert space}

The discussion so far has used (in contrast to \cite{Zurek07}, where pure states and scalar products sufficed to arrive at orthogonality) density matrices and Hilbert-Schmidt norm. This is certainly ``legal'', but it is possible to repeat derivation of quantum jumps using a purification procedure. Strictly speaking, this is unnecessary (after all, density matrices are a well-established way of representing states) but, as we shall see, purification yields new perspective on quantum jumps and actionability. Additional motivation stems from the use of quantum jumps in the derivation \cite{Zurek03a}
of Born's rule \cite{Born}, where they define ``events''. Emergence of events from unitarity and repeatability sets the stage for probabilities. It is therefore important to carry out the discussion that motivates probabilities without appealing to Born's rule. We have adhered to this goal above, and we have used density matrices only as mathematical tools (e.g., never appealing to their probabilistic interpretation). It is nevertheless reassuring that the proof can be repeated using only pure states: This should put to rest any suspicions about the possible circularity in motivating events -- a prelude to the envariant derivation of Born's rule.

Our strategy is simple: Mixed state of $\cS$ can be always represented by a Schmidt decomposition of a pure state:
$$ \ket {\Gamma_u^{\cS \cS'}} = \sum_k s_k \ket {\sigma_k^u} \ket {\sigma_k'} $$
The density matrix of $\cS$ is then
$\rho^u_{\cS} = \Tr_{\cS'} \kb {\Gamma_u^{\cS \cS'}}{\Gamma_u^{\cS \cS'}}$.
Consider now a unitary interaction of $\cS$ prepared in two states $\ket {\Gamma_u^{\cS \cS'}}$ and $\ket {\Gamma_v^{\cS \cS'}}$ with $\cA$:
$$\ket {\Gamma_u^{\cS \cS'}} \ket {A_0} \ket {A_0'}...\rightarrow \ket {\Gamma_u^{\cS \cS'}} \ket {A_u} \ket {A_0'}...\rightarrow \ket {\Gamma_u^{\cS \cS'}} \ket {A_u} \ket {A_u'}...\eqno(11u)$$
\vspace{-0.7cm}
$$\ket {\Gamma_v^{\cS \cS'}} \ket {A_0} \ket {A_0'}...\rightarrow \ket {\Gamma_v^{\cS \cS'}} \ket {A_v} \ket {A_0'}...\rightarrow \ket {\Gamma_v^{\cS \cS'}} \ket {A_v} \ket {A_v'}...\eqno(11v)$$
Repeatability implies that:
$$ \bk {\Gamma_u^{\cS \cS'}}{\Gamma_v^{\cS \cS'}}=\bk{\tilde \Gamma_u^{\cS \cS'}}{\tilde \Gamma_v^{\cS \cS'}} \bk {A_u} {A_v} $$ 
\vspace{-0.7cm}
$$ \bk {\tilde \Gamma_u^{\cS \cS'}}{\tilde \Gamma_v^{\cS \cS'}}=\bk{\tilde{\tilde \Gamma}_u^{\cS \cS'}}{\tilde{\tilde \Gamma}_v^{\cS \cS'}} \bk {A_u'} {A_v'} $$
As before, unitarity means that scalar products are preserved. Hence, we are now led to conclude that $\cA$ (or $\cA'$) can acquire distinct states that register even partial differences between $ \ket {\Gamma_u^{\cS \cS'}}$ and $\ket {\Gamma_v^{\cS \cS'}}$ only when:
$$ \bk {\Gamma_u^{\cS \cS'}} {\Gamma_v^{\cS \cS'}} = \sum_k |s_k|^2 \bk {\sigma_k^u}{\sigma_k^v}=0 \eqno(12)$$
Above, we recognized that the purifier $\cS'$ was unaffected by the interaction of $\cS$ and $\cA$, so that $\bk {\sigma_k'}{\sigma_l'} = \delta_{kl}$.

The above condition is also the promised example of actionable information that is non-local. Equation (12) differs from the simpler demand that the states of $\cS$ which can be repeatably copied should be orthogonal. It is of course satisfied when $\bk {\sigma_k^u}{\sigma_k^v} = \delta_{uv} \ \forall_k$, but this is not necessary:
Scalar products $\bk {\sigma_k^u}{\sigma_k^v}$ can have complex values or even just alternating signs such that while individual terms differ from zero, their sum in Eq. (12) adds up to 0. Bell states offer a simple example:
$$\ket { \gamma_{\pm}^{\cS \cS'} }= (\ket 0_\cS \ket 0_{\cS'} \pm \ket 1_\cS \ket 1_{\cS'})/\sqrt 2$$
For both of these states reduced density matrices are the same, $\rho^{\cS}_+=\rho^{\cS}_-\propto \bf 1$, so it is obviously impossible for $\cA$ that interacts only with $\cS$ to detect any difference between them. There can be no {\it local} actionable information in the state of $\cS$. Yet, in accord with Eqs. (2) and (11);
$$ \ket { \gamma_{\pm}^{\cS \cS'} } \ket {A_0} \rightarrow  \ket { \gamma_{\pm} } \ket {A_{\pm}} $$
is allowed, as $\bk { \gamma_{\pm}^{\cS \cS'}} { \gamma_{\mp}^{\cS \cS'}}=0$. As in Eq. (12), orthogonality is enforced on the composite state in $\cH_{\cS\cS'}$. $\cA$ acquires information about global property of $\cS\cS'$, relative phase distinguishing the two Bell states. Local interactions of $\cA$ with $\cS$ alone cannot be used to find out (copy) phases of Schmidt coefficients as can be seen using envariance \cite{Zurek03a}, 
Global phases have no local consequences. Thus, accessing global phases is possible only through global interactions. 

\section{Summary and Conclusions}

This paper generalizes the derivation of the discreteness underlying quantum jumps \cite{Zurek07} to macroscopic quantum system, such as a measurement apparatus. This is an important generalization -- it allows one to explore the consequences of repeatability in the macroscopic domain, where it is relevant for the quantum-to-classical transition. 
Macroscopic apparatus is open (and, hence, as other macroscopic systems, it can decohere), so distinct records correspond to mixed states. Repeatability does not require preservation of microscopic states -- they can be perturbed as long as their salient properties that determine the record inscribed in the state of the copies are preserved. 

Other interesting generalizations include \cite{ZP,Luo,LuoWei,Herb}. In particular, work of Luo and Wei \cite{LuoWei} suggests a connection of quantum jumps and no-broadcasting theorems \cite{Barnum,Lindblad,Piani}. We have not used no-broadcasting theorems because coarse-graining is key for the problem we have formulated. Thus, instead of replicating a complete mixed state we are abstracting only some aspects of that state to determine the states of the copies. 
Such abstracting of a part of the state changes the nature of the problem, and the original no-broadcasting theorem \cite{Barnum,Lindblad} is no longer directly applicable. It may be however possible that one could use theorems about broadcasting correlations \cite{Piani, LuoWei} to discuss ``quantum jumps''. 

Orthogonality of distinct sets of original states that correspond to e.g. different records in the apparatus (now represented by whole subspaces of the Hilbert space) parallels orthogonality of pure states in \cite{Zurek07}. Both follow from the core quantum postulates -- superposition principle, unitarity, and repeatability. In both cases discreteness of quantum jumps -- key symptom of collapse -- is a consequence of the core quantum postulates, and (contrary to the standard textbook lore) does not need to be postulated by additional axioms. Our derivation confirms Bohr's intuition about the role of information transfer, and yet it relied only on unitary evolutions. As is the case for decoherence this derivation of the symptoms of quantum jumps and the wavepacket collapse can be adapted to fit either post-Copenhagen or post-Relative States point of view.


This research was supported in part by DoE through LDRD grant at Los Alamos, and in part by the John Templeton Foundation. Stimulating discussions with Jess Riedel and Alex Streltsov are greatly appreciated.


\begin{thebibliography}{99}
\bibitem{foot} One could also consider degeneracies in the implementation of actionability. For instance, the measurement distinguishing $u$ from $v$ could be just one datum in a more complex algorithm where several possible output states $\tau_{u1}, \tau_{u2}, ... $ and $\tau_{v1}, \tau_{v2}, ...$ may appear on the output. Such situations may be of interest in quantum information processing, but their discussion is beyond the scope of our paper which is focused on orthogonality.
\bibitem{note} Of course, when there exists unitary ${\cal U}_{{\cA^{(k)}}{\cal T}}$ that can conditionally transform mixed state $\tau_0$ of $\cal T$ into states that distinguish $\rho_{u\cA^{(k)}} $ from $\rho_{v\cA^{(k)}} $,  one can always find pure states of $\cal T$ that will be similarly affected.
\end{thebibliography}
\end{document}